\documentclass[runningheads]{llncs}
\usepackage[T1]{fontenc}
\usepackage{graphicx}
\usepackage{booktabs}
\usepackage{array}
\usepackage{tabularray}

\usepackage{amsfonts}

\usepackage{amsmath}

\usepackage{subfig}
\usepackage{hyperref}
\usepackage{svg}

\begin{document}
%
%\title{Generalized Encrypted Traffic Classification: Multi-Level Analysis with Inter-Flow Signals}
\title{Generalized Encrypted Traffic Classification Using Inter-Flow Signals\thanks{This work was supported by project SERICS (PE00000014) under the
NRRP MUR program funded by the EU-NGEU.}}
%Multi-Task Encrypted Traffic Classification through Inter-Flow Signals
%Leveraging Inter-Flow singnals for Generalized Encrypted Traffic Classification

\author{Federica Bianchi\inst{1}\orcidID{0009-0006-2698-4484} \and
Edoardo Di Paolo\inst{1}\orcidID{0000-0001-9216-8430} \and
Angelo Spognardi\inst{1}\orcidID{0000-0001-6935-0701}}
% %
\authorrunning{Bianchi et al.}
\institute{Computer Science Department, Sapienza University of Rome,\\
      % viale Regina Elena 295, Rome, Italy\\
    \email{bianchi@di.uniroma1.it},
    \email{dipaolo@di.uniroma1.it},
    \email{spognardi@di.uniroma1.it}
}
\maketitle              % typeset the header of the contribution
\begin{abstract}
In this paper, we present a novel encrypted traffic classification model that operates directly on raw PCAP data without requiring prior assumptions about traffic type. Unlike existing methods, it is generalizable across multiple classification tasks and leverages inter-flow signals—an innovative representation that captures temporal correlations and packet volume distributions across flows. Experimental results show that our model outperforms well-established methods in nearly every classification task and across most datasets, achieving up to 99\% accuracy in some cases, demonstrating its robustness and adaptability.

\end{abstract}
\section{Introduction}
Network traffic analysis is a fundamental aspect of cybersecurity, network management, and performance optimization. Analyzing network activity provides valuable insights into communication patterns, the services and applications in use, and potential security threats. As a result, traffic analysis plays a key role in a wide range of real-world applications, from securing enterprise networks and detecting malware to enhancing service quality and protecting user privacy.
In recent years, the increasing adoption of encryption protocols has rendered plaintext payload analysis ineffective, shifting the focus to methods based on machine learning and deep learning. 
Encrypted traffic analysis supports multiple critical goals~\cite{shen2023survey}, including the \textit{identification of network assets} like IoT systems and mobile devices, enabling better asset management and vulnerability assessment. It also facilitates \textit{network characterization} by evaluating service quality and user experience metrics. Moreover, it plays a key role in \textit{privacy leakage detection}, revealing user activities and accessed apps, services, or websites through encrypted traffic. Finally, it is crucial for identifying malicious behavior and anomalies, a need underscored by the rising prevalence of malware and attacks on enterprise, IoT, and blockchain infrastructures.

Despite the broad applicability of encrypted traffic classification, most existing research typically focuses on developing a method tailored to a single task and is rarely evaluated in multiple analysis goals, limiting the generalizability.
Moreover, previous work focuses mainly on statistical features extracted from packets or within individual flows (intra-flow level), with only a few exploring relationships among multiple flows over time (inter-flow level) and integrating insights from multiple perspectives~\cite{pham2021mappgraph,vanede2020flowprint,wang2024interflow_spectograms}.

To address these limitations, we present a novel encrypted traffic classification model that is both generalizable across multiple domains and does not require prior assumptions about traffic type. Our approach proves effective across various tasks, including mobile application and website fingerprinting, malware detection, IoT device fingerprinting, and general traffic classification (with and without VPN), making it a highly versatile solution. Although our model has been tested on the aforementioned tasks, it also holds significant potential for application in other encrypted traffic classification tasks due to its generalizability. Another key innovation of our method is the introduction of signals, a novel inter-flow representation that captures temporal correlations among flows and packet volume distributions. 
Drawing inspiration from telecommunications and signal processing~\cite{picone1993speech}, we treat network flows as temporal signals. Unlike statistical features, which summarize flow characteristics in isolation, signals model how multiple flows interact over time, offering a richer context-aware classification of encrypted traffic.
To our knowledge, this is the first general-purpose encrypted traffic classifier using inter-flow signal representations.

The main contributions of this paper can be summarized as follows:

\begin{itemize}
    \item We propose a general-purpose encrypted traffic classification model that successfully generalizes across multiple analysis goals.
    \item We introduce a novel inter-flow representation, called signals, which captures temporal correlations among flows and packet volume distributions.
    \item We conduct an extensive evaluation across eight diverse datasets, covering multiple encrypted traffic classification tasks. 
    \item We compare our approach against well-established methods across the same datasets, demonstrating superior performance in all classification tasks.
\end{itemize}

\section{Related Work}
\label{section:related_work}

One of the main applications of encrypted network traffic analysis has been the identification of mobile applications.
FlowPrint~\cite{vanede2020flowprint} analyzed temporal correlations among destination-related features of network traffic at the inter-flow level and created a database of application-specific fingerprints. When classifying new traffic, their method generated a corresponding fingerprint and matched it against the database. %Other knowledge-based methods use DNS correlation~\cite{knowledge_He} to identify apps or manually identify specific user actions~\cite{knowledge_sequential}.
\textit{Taylor et al.} in AppScanner~\cite{taylor2018appscanner} adopted machine learning models using statistical features at the intra-flow level and packet sizes to recognize apps.
More recent methods leverage deep learning. \textit{Pham et al.}~\cite{pham2021mappgraph} proposed MAppGraph, a method that classifies mobile applications by constructing a communication graph of network destinations for each app. The nodes of the graph for each app are defined by tuples of IP addresses and ports of the services connected by the app, while the edges represent weighted communication correlations among these nodes.
They then used deep graph convolutional neural networks (DGCNN) to learn the various communication behaviors of mobile apps.
Some works, instead, feed raw packet bytes directly into deep learning models, such as in PEAN~\cite{lin2023pean}. 
 
Website fingerprinting aims to identify the specific website visited by a user despite encryption mechanisms. 
Many methods leverage machine learning to infer web activity~\cite{shen2020optimizing_feature,shen2019secondaryencryption}, while others leverage deep learning approaches, such as in~\cite{cui2020realistic_website_fing,zhang2020datadriven}, in which they classify websites visited using CNNs. 

Generic traffic classification targets broad service categories (e.g., email, VoIP).
\textit{Chen et al.}~\cite{chen2021sequential_message} employed Long Short-Term Memory (LSTM) on message size sequences for early prediction, while \textit{Luo et al.}~\cite{luo2024flow_embedding} proposed a self-supervised method for traffic classification to enhance flow representations using trace data.

Encrypted traffic analysis has also been widely used to distinguish malicious activity from benign traffic. The approaches range from traditional machine learning classifiers~\cite{garg2017networkadroidmalicious,lashkari2017android_malware} to deep models.
\textit{Feng et al.}~\cite{feng2020twolayermalware} introduced a two-layer CNN-AutoEncoder to classify malware, while more recently \textit{Liu et al.}~\cite{liu2024ltachecker} modeled temporal patterns in Dalvik opcode sequences with LSTM and temporal convolutional network (TCN).
In IoT device fingerprinting, statistical profiling has been effective~\cite{sivanathan2017characterizingiot}, while recent efforts employ deep learning for IoT malware detection~\cite{ali2023iotmalwaredetection}.

\section{Problem Definition}
\label{sec:problem_definition}
Encrypted network traffic classification categorizes network traffic according to different analysis goals. Indeed, traffic can be generated from various sources, such as mobile applications, web pages or websites, services, malware, etc.

Unlike approaches that classify individual traffic \textit{flows} or \textit{packets}, our method operates on \textit{traffic chunks}. A traffic chunk is a bounded segment of network traffic that spans a specific time window, which may include multiple concurrent or sequential flows.   
Formally, let \( P \) be the set of all the observed\textbf{ \textbf{network packets}} in a monitored traffic session. Each packet \( p_j \in P \) belongs to a specific \textbf{network flow}, which consists of a series of packets sharing the standard 5-tuple:
\begin{equation}
\langle \textit{src}_{\textit{ip}}, \textit{dst}_{\textit{ip}}, \textit{src}_{\textit{port}}, \textit{dst}_{\textit{port}}, \textit{proto} \rangle,
\label{eq:flow_tuple}
\end{equation}

where \( \textit{src}_{\textit{ip}} \) and \( \textit{dst}_{\textit{ip}} \) are the source and destination IP addresses, \( \textit{src}_{\textit{port}} \) and \( dst_{\textit{port}} \) are the source and destination ports, and \textit{proto} is the transport-layer protocol.
A \textbf{traffic chunk} \( C_i \) is defined as a set of packets occurring within a fixed time window \( W \), such that:
\begin{equation}
C_i = \{ p_j \in P \mid t_j \in [t_{\textit{start}}, t_{\textit{start}} + W] \}.
\end{equation}
where \( t_{j} \) is the timestamp of packet \( p_{j} \), \( t_{\textit{start}} \) is the starting time of the chunk and \( W \) is the pre-defined window duration. 

A traffic chunk may contain different flow communication scenarios:
\begin{itemize}
    \item \textbf{Single Flow Communication}: Occurs when a network interaction involves one isolated flow between a client and server, exhibiting simple request-response behavior, without concurrent or sequential connections. %This is typically observed in protocols with persistent connections, where multiple packets are exchanged back and forth without initiating new connections. 
    \item \textbf{Sequential Flow Communication}: Involves multiple flows initiated sequentially for different stages of the communication. Each flow contributes to a broader activity.%, such as during malware infections, where an initial beaconing flow to a C2 server is followed by flows retrieving additional payloads. 
    \item \textbf{Concurrent Flow Communication}: In many real-world scenarios, endpoints initiate multiple flows in parallel towards different network destinations to manage different aspects of the communication. 
\end{itemize}

Since we evaluate our model on multiple classification tasks, the ground-truth label varies depending on the task. For mobile applications, it identifies the specific application in use. In website classification, it corresponds to the accessed site, whether through standard or VPN-protected browsing. For malware detection, the label distinguishes between benign and malicious traffic. In service classification, it indicates the type of network service (e.g., chat, email, VoIP, etc.). Finally, for IoT classification, it reflects the type of IoT device involved in the communication.
\section{Methodology}
\label{section:approach}

We propose a general traffic classification method that operates directly on raw PCAP files, without assumptions about the nature of the traffic.
This flexible design supports various analysis goals, including mobile application fingerprinting, website fingerprinting, generic traffic classification, malware detection, and IoT traffic classification.
To achieve effective classification, we extract features at multiple representation levels. At the inter-flow level, we introduce \textbf{signals}, a key novelty of our work, which capture temporal relationships, volume, and timing across flows within the same time window. We also compute statistical features at the packet and intra-flow levels for detailed insights into flow behavior and packet distributions.

\textbf{Traffic Pre-Processing.}
Following prior work~\cite{oh2023appsniffer,pham2021mappgraph,vanede2020flowprint}, we filter out packets based on port numbers, discarding traffic from common background services (e.g., DNS, DHCP). These ports are typically not related to specific apps, services, or websites and do not offer distinguishing characteristics for classification~\cite{pham2021mappgraph,zhang2020datadriven}. Removing them reduces irrelevant variation and improves the clarity of traffic features for classification.
After PCAPs are filtered, we split them into smaller \textbf{chunks}, or time windows, of network activity, to improve computational efficiency and analytical precision.
Segmenting traffic into manageable portions reduces the complexity and overhead associated with analyzing large captures, essential for real-time or high-volume scenarios.
Additionally, breaking down traffic into time windows allows us to focus on discrete periods of network communication, enabling a finer-grained analysis of the traffic temporal dynamics, as network behavior can vary over time.
To prevent information loss at the boundaries of these time windows, we introduce an \textbf{overlap} parameter, which defines the fraction of two consecutive chunks that overlap. Without overlap, some traffic patterns occurring near chunk edges may be split between consecutive windows, leading to fragmented representations.
Choosing the time window duration and overlap involves trade-offs. Short windows capture fine-grained interactions but can be noisy and fragmented, while longer ones smooth out fluctuations and provide a broader perspective, but risk obscuring short-term behaviors and delaying classification. Similarly, large overlaps preserve continuity but add redundancy, whereas small overlaps are more efficient but may disrupt dependencies between related flows. We empirically tune these parameters for optimal balance. \\
\textbf{Traffic Representation.}
To achieve a thorough understanding of traffic patterns, we represent them at multiple levels: packet level, intra-flow level, and inter-flow level.
This choice comes from the intuition that traffic may leak different kinds of information based on the viewpoint of the analysis, each providing valuable insights into its behavior and the overall dynamics of the network. 
%\textbf{Signals as Inter-Flow Features.}
At the \textbf{inter-flow level}, we aim to leverage the \textbf{temporal relationships} between the sending and receiving of packets across multiple flows. The key hypothesis underlying this representation is that network communications generate discernible patterns in flow initiation and sequencing, which can be used for traffic classification.
When analyzing the network behavior of communicating endpoints within specific time windows (i.e., chunks), we observe the scenarios outlined in Section~\ref{sec:problem_definition}, namely single flow communication, sequential flow communication, and concurrent flow communication.
Traditional studies on encrypted traffic classification typically isolate individual flows for analysis. In contrast, our approach wants to capture these sequential and concurrent flow communication scenarios, examining how flows \textbf{coexist} and \textbf{interact} within the same temporal context. This provides a more holistic representation that integrates both the byte volume and the temporal relationships between flows, such as flows initiation, sequencing, and overlap. 
%These relationships are valuable in distinguishing between different types of traffic.
To capture these inter-flow dynamics, we represent network traffic as a discrete time series signal. This transformation moves beyond simple flow classification by analyzing traffic as a structured sequence that encapsulates both the exchanged data volume and the key temporal features of flow interactions.
The general idea is that each chunk is transformed into a \textbf{unified signal} that aggregates all packets exchanged during that time window, regardless of whether they originate from a single flow or multiple concurrent flows. The rationale behind constructing a unified signal is to capture the endpoint global communication pattern, rather than focusing on individual flows. \\
With this unified signal, we aim to: (i) capture the volume of transmitted data across all flows within the chunk, along with the temporal structure of packet transmissions, revealing distinct traffic patterns; (ii) quantify each flow contribution to the overall network activity, helping to identify dominant interactions in data transfer; and (iii) analyze flow interactions to help distinguish different types of traffic. For example, mobile applications may exhibit consistent flow initiation and overlap patterns---such as connections to authentication servers, content delivery networks, or analytics services. This predictable behavior suggests that the initiation of flows follows a relatively consistent pattern in both timing and volume.

\paragraph{Signal Creation.} Each network packet is represented as a tuple containing its \textbf{timestamp} and \textbf{packet length}:
\begin{equation}
    p_j = (t_j, l_j),
\end{equation}
where \( t_j \) represents the time at which the packet was transmitted as recorded in the PCAP file (approximated in seconds for analysis) and \( l_j \) denotes its size in bytes. A network flow \( F_k \) is then defined as an ordered sequence of packets that share the same five-tuple identifier, as defined in \autoref{eq:flow_tuple}:
\begin{equation}
    F_k = \{ (t_1^k, l_1^k), (t_2^k, l_2^k), ..., (t_{n_{k}}^k, l_{n_{k}}^k) \} \quad \text{where } t_1^k < t_2^k < \dots < t_{n_{k}}^k.
\end{equation}
Each flow consists of packets exchanged between a client and a server, and multiple flows may be present within a given traffic chunk, as described in the previous sections.
To construct a single \textbf{unified signal} from multiple flows (sequential or concurrent), we aggregate their packet information over a coherent time axis to ensure that all flows are synchronized within the time window of the chunk, aligning their packet data into a cohesive representation.

Let \( F = \{F_1, F_2, ..., F_n\} \) represent the set of flows within a chunk. We determine the minimum and maximum timestamps across all flows:
\begin{equation}
    t_{\text{min}} = \min_{k=1,\cdots, n} \min_{i=1,\cdots, n_{k}} t_i^{k} , \quad 
    t_{\text{max}} = \max_{k=1,\cdots, n} \max_{i=1,\cdots, n_{k}} t_i^{k}.
\end{equation}
Then, we define \(T\) as a discretization of the interval \(t_{\text{min}} \text{ to } t_{\text{max}} \) with a step size \( \delta \), which can be chosen arbitrarily:
\begin{equation}
    T = \{ t_{\text{min}}, t_{\text{min}} + \delta, ..., t_{\text{max}} \}.
\end{equation}
For each flow \( F_k \), we compute its \textbf{amplitude}, which represents the total transmitted data volume, computed as follows:
\begin{equation}
    A_k = \sum\limits_{j=1}^{n_{k}} l_j^k.
\end{equation}
Each packet is then mapped onto the time axis, and the value of the signal at each timestamp is computed by summing the contributions of all active flows:
\begin{equation}
    S(T_{min} + \delta t) = \sum\limits_{k = 1}^n A_{k} \sum\limits_{j = 1}^{n_{k}} l_j^k \mathbf{1}_{[T_{min} + \delta t, T_{min} + \delta (t+1))} (t_{j}^k),
    \quad
    t = 0, \cdots, \frac{T_{max} - T_{min}}{\delta}.
\end{equation}
where, given a set B,
\begin{equation}
    \mathbf{1}_{B}(x) = \begin{cases}
        1 \quad & \text{if } x \in B, \\ 0 \quad & \text{otherwise.} 
    \end{cases}
\end{equation}
By scaling the signal values with amplitude, we ensure that each flow influence is proportional to its total traffic volume within the chunk. High-volume flows contribute more significantly, while low-volume flows have a lesser impact.

\begin{table}[t!]
\centering
\caption{List of features extracted for incoming packets, outgoing packets, and their combined set.}
\label{tab:packet_intra_flow_features}
\resizebox{\linewidth}{!}{%
\begin{tblr}{
  rows = {m},
  hline{2-3} = {-}{},
}
 & \textbf{Features}\\
\textbf{\textbf{Packet-Level}} & {Total Number of Packets, Maximum Packet Size, Minimum Packet Size, \\Mean Packet Size, Variance of Packet Size, Standard Deviation of Packet Size,\\Mean Absolute Deviation of Packet Size, Skewness of Packet Size,\\~Kurtosis of Packet, Size Percentiles (10th to 90th)}\\
\textbf{\textbf{Intra-Flow Level}} & {Mean Flow Duration (Seconds), Mean Flow Size (Bytes), Total Number of Flows,\\Standard Deviation of Flow Duration, Mean Number of Packets per Flow}
\end{tblr}
}
\end{table}

Min-Max Normalization is applied to maintain comparability across signals of different chunks.
In addition to constructing the signal, we compute and normalize the inter-arrival times of packets across all flows.
These inter-arrival times capture patterns such as burstiness, network congestion, or application-specific timing characteristics.
By incorporating both the aggregated traffic signal and inter-arrival times, our method effectively captures the overall data volume and flow significance and the timing characteristics of packet exchanges. \\
To our knowledge, this approach of deriving a single application-level signal from overlapping flows has not been previously explored in the literature.

At the \textbf{packet} and \textbf{intra-flow levels}, we extract statistical features that describe both individual packets and their corresponding flows. This choice is motivated by their demonstrated effectiveness in the literature~\cite{shen2020optimizing_feature}, where statistical features have consistently yielded strong results in network traffic analysis. 
In particular, packet level features capture the statistical properties of individual packets within a chunk. These features are extracted separately for incoming packets, outgoing packets, and their combined set. 
On the other hand, intra-flow features describe the behavior of individual flows, capturing how packets are transmitted over time. %These features are crucial in understanding how endpoints communicate over time and how long-lived sessions are structured. 
All the extracted features are summarized in \autoref{tab:packet_intra_flow_features}. \\
\textbf{Traffic Analysis.}
During Traffic Analysis, the extracted features are processed to derive meaningful insights about the encrypted network traffic. 
Our approach derives insights from traffic patterns and distributions that describe the behavior of network flows and packets. 
We employ \textbf{Random Forest}, a well-established machine learning technique effective in high-dimensional data, such as that from network traffic.

\section{Evaluation}
\label{section:evaluation}
We implemented our model in Python, using Scapy \cite{scapy} PCAP processing and analysis and the scikit-learn library for the Random Forest classifier.
To assess the performance of our proposed method, we conducted an evaluation on eight different datasets (MAppGraph~\cite{pham2021mappgraph}, PostQuantumTLS~\cite{mankowski2023postquantum_tls_dataset}, Cross Platform~\cite{ren2019crossplatfrom}, ISCX-VPN and ISCX-nonVPN~\cite{lashkari2016iscx}, CICAndMal2017~\cite{lashkari2018cicandmal2017}, IoT-Sentinel~\cite{miettinen2017iotsentinel} and CSTNET-TLS 1.3~\cite{lin2022bert}) and compared our approach with Flowprint \cite{vanede2020flowprint} and MAppGraph \cite{pham2021mappgraph}.
Flowprint was selected as it is a well-established knowledge-based approach for mobile traffic classification, while MAppGraph is one of the few methods that classify chunks of traffic, making it a suitable choice for direct comparison with our approach.

%\subsection{Experimental Setup}
We evaluated our model on various datasets in order to address an assorted range of network traffic classification challenges, including mobile application fingerprinting, website fingerprinting, traffic classification (with and without VPN), and IoT device fingerprinting.
All experiments use an 80:20 train-test split and standard metrics—accuracy, precision, recall, and F1-score. 
We tested various window-overlap configurations to optimize performance per dataset. The evaluated time window sizes included \textit{300s}, \textit{200s}, \textit{80s}, \textit{50s}, \textit{30s}, and \textit{10s}, with overlaps of \textit{180s}, \textit{120s}, \textit{30s}, \textit{10s}, \textit{2s}, and \textit{1s}. 
For each dataset, we selected the window-overlap combination that produced the highest accuracy, and all reported results correspond to this best-performing configuration.
For fair comparative evaluation, we used the official implementations of MAppGraph\footnote{\href{https://github.com/soeai/mappgraph.git}{https://github.com/soeai/mappgraph.git}} and FlowPrint\footnote{\href{https://flowprint.readthedocs.io/en/latest/index.html}{https://flowprint.readthedocs.io/en/latest/index.html}}. 
The MAppGraph implementation was used without modification, except for an added module to convert PCAP files to CSVs (as the original implementation assumes CSV input). We tested the same range of window sizes and overlaps for parameter tuning. For FlowPrint, we installed the library via \texttt{pip}, preprocessed the PCAPs to extract flows, generated fingerprints, and subsequently performed app recognition.

\begin{table}[t!]
%\vspace{-2.5em}
\centering
\caption{Results and comparison of our method with MAppGraph and Flowprint across different datasets.}
\label{tab:results_comparison}
\resizebox{\linewidth}{!}{%
\begin{tblr}{
  column{even} = {c},
  column{3} = {c},
  column{5} = {c},
  column{7} = {c},
  column{9} = {c},
  column{11} = {c},
  column{13} = {c},
  cell{1}{2} = {c=4}{},
  cell{1}{6} = {c=4}{},
  cell{1}{10} = {c=4}{},
  vline{2-3,7} = {1}{},
  vline{2,6,10} = {2-10}{},
  hline{2,3,6,8-10} = {-}{},
}
\textbf{Model} & \textbf{Our Method} &  &  &  & \textbf{MAppGraph} &  &  &  & \textbf{Flowprint} &  &  & \\
\textbf{Dataset} & \textbf{Acc.} & \textbf{Prec.} & \textbf{Rec.} & \textbf{F1} & \textbf{Acc.} & \textbf{Prec.} & \textbf{Rec.} & \textbf{F1} & \textbf{Acc.} & \textbf{Prec.} & \textbf{Rec.} & \textbf{F1}\\
\textbf{MAppGraph} & \textbf{0.9517} & 0.9473~ & 0.9385 & 0.9418 & 0.9346 & 0.9364 & 0.9346 & 0.9347~ & 0.8664 & 0.8718 & 0.8664 & 0.8662 \\
\textbf{PostQuantumTLS} & 0.5013 & 0.5624 & 0.5013 & 0.4941 & 0.3684 & 0.2663 & 0.2760 & 0.2612 & \textbf{0.7077} & 0.7177 & 0.7077 & 0.6927\\
\textbf{Cross Platform} & 0.3418 & 0.5088 & 0.3120 & 0.2748 & 0.2686 & 0.1853 & 0.2130 & 0.1862 & \textbf{0.8692} & 0.9070 & 0.8692 & 0.8745\\
\textbf{ISCX-VPN} & \textbf{0.9903} & 0.9895 & 0.9882 & 0.9885 & 0.8846 & 0.6988 & 0.7195~ & 0.7055 & 0.9484 & 0.9728 & 0.9484 & 0.9557\\
\textbf{ISCX-nonVPN} & \textbf{0.9918} & 0.9887 & 0.9811 & 0.9845 & 0.9244 & 0.9022 & 0.8875 & 0.8915 & 0.7791 & 0.8770 & 0.7791 & 0.8150\\
\textbf{\textbf{CICAndMal2017}} & 0.8308 & \textbf{ 0.8315} & 0.8285 & 0.8295 & \textbf{0.8518} & 0.7507 & 0.6467 & 0.6755 & 0.7836 & 0.7837 & 0.7836 & 0.7829\\
\textbf{IoT-Sentinel} & \textbf{0.9280} & 0.9198 & 0.9293 & 0.8878 & 0.6553 & 0.6002 & 0.5451 & 0.5481 & 0.7084 & 0.7084 & 0.7084 & 0.7084\\
\textbf{\textbf{CSTNET-TLS 1.3}} & \textbf{0.7497} & 0.7408 &  0.7079 &  0.7142 & - & - & - & - & 0.1953 & 0.1987 & 0.1953 & 0.1834
\end{tblr}
}
%\vspace{-2em}
\end{table}

\textbf{Analysis of Results.}
\label{subsec:performance}
Our method demonstrates (\autoref{tab:results_comparison}) robust classification performance on almost all tested datasets, achieving high accuracy across all analysis goals.
It operates without assumptions about the type of analysis goals considered and achieves consistent performance across a diverse range of datasets.
These results suggest that our signal-based inter-flow features, combined with statistical representations, effectively capture characteristics of different types of encrypted traffic.
In particular, our method achieves outstanding results in the general traffic classification task, reaching 99.03\% accuracy on ISCX-VPN and 99.18\% on ISCX-nonVPN.
Additionally, it performs very well in IoT device detection (92.80\% on IoT-Sentinel) and malware classification (83.08\% on CICAndMal2017).
However, we encountered limitations with datasets like Cross Platform and PostQuantumTLS.
These datasets pose a challenge, particularly for methods that classify chunks of traffic, as they contain only one PCAP per app, with a maximum capture duration of five minutes.
The short duration of these PCAPs results in very limited network activity, which complicates the extraction of meaningful statistical features.
When divided into smaller chunks, these limited captures result in segments that may lack the packet volume and traffic diversity needed to extract meaningful statistical features. 
Furthermore, signal generation can be affected by these short PCAPs, as the flow duration is limited and the number of concurrent flows captured is minimal. \\ 
\textbf{Comparison with MAppGraph and Flowprint.}
\autoref{tab:results_comparison} shows that our method outperforms existing approaches in most cases.
MAppGraph, like our method, classifies chunks of traffic rather than individual flows. It struggles with smaller datasets, such as CrossPlatform (26.96\%) and PostQuantumTLS (36.84\%), likely for our same reasons.
Despite sharing a chunk-based approach, MAppGraph’s performance remains lower than ours across most datasets. Notably, our method outperforms MAppGraph even on its own dataset, and it achieves higher accuracy in the general traffic classification task (ISCX-VPN and ISCX-nonVPN), website fingerprinting (CSTNET-TLS 1.3), and IoT device recognition (IoT-Sentinel), demonstrating its better adaptability across different traffic classification tasks. 
Although MAppGraph achieves a higher accuracy (85.18\%) compared to our method (83.08\%) on the CICAndMal2017 dataset, it demonstrates significantly lower precision (75.07\% vs. 83.15\%) and recall (64.67\% vs. 82.85\%). This distinction is important in malware detection tasks: lower precision means that MAppGraph is more likely to generate false positives, flagging legitimate traffic as malicious, while lower recall implies that MAppGraph may miss some malicious traffic, leaving malware undetected. 
FlowPrint, in contrast, uses a knowledge-based approach, generating fingerprints for applications.
This methodological difference may explain why FlowPrint performs better on smaller datasets, namely Cross Platform and PostQuantumTLS. However, despite its strong general performance across multiple classification tasks, FlowPrint never surpasses our method except for the previously mentioned datasets.
Importantly, both FlowPrint and MAppGraph struggle in the CSTNET-TLS 1.3 dataset. This is likely due to their reliance on network destinations and ports for graph construction. When working with websites, especially those with fewer unique IP-port pairs, these methods face difficulty in creating meaningful graphs and in capturing enough network diversity to make accurate classifications. 
While these approaches work well for mobile application fingerprinting, for which they were specifically developed—due to the typical presence of many distinct network destinations contacted by apps—they do not perform as well for websites. As a result, on CSTNET-TLS 1.3 dataset, MAppGraph is unable to even create the graphs needed to test the model, while FlowPrint achieves only 19.53\% of accuracy. This limitation highlights a failure in generalizability.

\section{Conclusions and Future Works}
\label{section:conclusions}

In this paper, we proposed a general-purpose encrypted traffic classification model that operates across multiple analysis goals without prior assumptions about traffic type. A key innovation of our approach is the signal-based inter-flow representation, which captures temporal correlations and packet volume distributions across flows, providing a richer and more context-aware representation of encrypted network traffic.
Our evaluation across eight datasets demonstrates the effectiveness and adaptability of our method, consistently outperforming state-of-the-art models in mobile application recognition, website fingerprinting, malware detection, IoT fingerprinting, and general traffic classification. 
Future work will focus on improving performance with small datasets and short PCAPs, exploring with other machine and deep learning models for enhanced accuracy, and investigating real-time encrypted traffic detection for dynamic environments.

\begin{credits}
%\subsubsection{\ackname} A bold run-in heading in small font size at the end of the paper is..
% used for general acknowledgments, for example: This study was funded

% \subsubsection{\discintname}
% It is now necessary to declare any competing interests or to specifically
% state that the authors have no competing interests. Please place the
% statement with a bold run-in heading in small font size beneath the
% (optional) acknowledgments\footnote{If EquinOCS, our proceedings submission
% system, is used, then the disclaimer can be provided directly in the system.},
% for example: The authors have no competing interests to declare that are
% relevant to the content of this article. Or: Author A has received research
% grants from Company W. Author B has received a speaker honorarium from
% Company X and owns stock in Company Y. Author C is a member of committee Z.
\end{credits}
\bibliographystyle{splncs04}
%\bibliography{bibliography}

\end{document}